\documentclass[12pt]{iopart}
\usepackage{graphicx}

\begin{document}

\title[From non-Hermitian effective operators to the no-core shell model]
{From non-Hermitian effective operators to large-scale no-core
shell model calculations for light nuclei}

\author{Bruce R. Barrett$^1$, Ionel Stetcu$^1$\footnote{On leave from the 
National Institute for Physics and Nuclear Engineering ``Horia Hulubei",
Bucharest, Romania.},
Petr Navr\'{a}til$^2$,
and James P. Vary$^{2,3}$}

\address{$^1$Department of Physics, University of Arizona, Tucson 85721 \\
$^2$Lawrence Livermore National Laboratory, Livermore, 
P.O. Box 808, California 94551\\
$^3$Department of Physics and Astronomy, Iowa State University, Ames,
Iowa 50011}

\begin{abstract}
No-core shell model (NCSM) calculations using \textit{ab initio} effective
interactions are very successful in reproducing experimental nuclear
spectra. The main theoretical approach is the use of effective operators,
which include correlations left out by the truncation of the model space
to a numerically tractable size. We review recent applications 
of the effective operator approach, within a NCSM framework,
to the renormalization of the nucleon-nucleon interaction, as well as
scalar and tensor operators.
\end{abstract}


\section{Introduction}

The theory of effective operators plays an important role in the modern
approach to nuclear structure. Effective interactions are the basic
ingredient of the no-core shell model (NCSM), one of the \textit{ab
initio} methods that provides solution to the nuclear many-body problem
starting from high precision nucleon-nucleon (NN) interactions (\textit{i.e.},
that describe the two-nucleon data with high accuracy) and theoretical
three-nucleon forces.

Numerical solution to the $A$--body Schr\"odinger equation can be obtained
only if one truncates the Hilbert space to a finite, yet sufficiently small
dimension. Restriction of the space to a numerically tractable size
requires that operators for physical observables be
replaced by effective operators that are designed
to account for such effects. Most applications of the effective
operator theory are limited to deriving effective interactions, but other
observables are of great interest as well. In particular, for electromagnetic
operators, a long standing problem in the phenomenological
shell model was the use of effective charges for protons and
neutrons. Perturbation theory has been partially successful in describing
empirical effective charges needed to explain experimental transition
strengths \cite{Osnes}. However, recent
investigations using the unitary
transformation approach within the framework of the NCSM to obtain
effective operators have reported some progress in explaining the 
large values of the empirical effective charges 
\cite{Navratil:1996jq}. We will discuss briefly this result and its 
consequences later.

In the restricted space, the effective operators are constructed
to reproduce the values of the corresponding physical 
observables in the full space.
However, the renormalization procedure usually alters properties
of bare operators; for example, the interaction is no longer Hermitian,
and general transition operators change their rotational symmetry
properties.
While in some cases non-Hermitian Hamiltonians have advantages
\cite{Bender:2005sc}, in our case this presents a major inconvenience. 
Moreover, some approaches introduce energy dependence of the
resulting effective operators, an additional complication for
solving the nuclear many-body problem. 
This drawback is, however, avoided in the unitary transformation
approach to effective operators of Okubo \cite{Okubo:1954} and others
\cite{DaProvidencia:1964,Suzuki:1980,Suzuki:1982,Suzuki:1994ok}.
This method allows us to construct all effective
operators in an energy independent form, and, through an additional
similarity transformation, to restore the Hermiticity of the
effective interaction and the roattional properties of transition operators.

This paper is organized as follows: we review the
theoretical approach in Sec. \ref{thAp},
and then apply the procedure in realistic cases, using realistic
two-body interactions,
for the Hamiltonian in Sec. \ref{effInt}, and other general operators in Sec.
\ref{genEffOp}. We draw our conclusions in Sec. \ref{concl}.

\section{Theoretical Approach}
\label{thAp}

In this section we review the similarity transformation approach to effective operators and
discuss its practical implementation in the case of the nuclear many-body Hamiltonian. 

\subsection{Formal theory}

It is not our intention to discuss in great detail the method; 
we will point out the main features, following the 
derivation in Refs. \cite{Navratil:1993plb} and \cite{Navratil:1993NPA}.

In our approach, the full Hilbert space is divided into a model space,
with associated projection operator $P$, and a complementary,
excluded space, with the associated projection operator $Q$ ($P+Q=1$).
The goal is to perform many-body calculations in the model space,
using a transformed Hamiltonian
$\mathcal{H}$,
\begin{equation}
{\mathcal{H}}=XHX^{-1},
\label{transfHam}
\end{equation}
so that a finite subset of eigenvalues of the initial Hamiltonian
$H$ are reproduced. We need to point out that this is a general
approach, which can be applied to non-Hermitian Hamiltonian operators 
that can arise, for example, in the context of boson mappings.

To better understand the conditions that we will impose on the 
transformation operator $X$, we start with the
results of the Feshbach projection formalism on the Schr\"odinger
equation
\begin{equation}
{\mathcal H}|\Psi\rangle=E_\Psi|\Psi\rangle.
\label{schr0}
\end{equation}
In general for non-Hermitian Hamiltonians, the left and
right eigenvectors are not related simply by a Hermitian conjugation, but we
have the freedom to choose a normalization so that 
$\langle\tilde\Psi_E|\Psi_E\rangle=1$, where $\langle \tilde\Psi_E|$ is
the left eigenvector corresponding to the eigenvalue $E_\Psi$.
It follows from Eq. (\ref{schr0}) that the 
component of the wave function $|\Psi\rangle$ outside the model space
is given by
\begin{equation}
Q|\Psi\rangle= \frac{1}{E_\Psi-Q{\mathcal H}Q}Q{\mathcal H}P|\Psi\rangle,
\label{QrightEV}
\end{equation}
so that the effective Hamiltonian in the model space can be expressed as
\begin{equation}
H_{eff}=P{\mathcal H}P+P{\mathcal H}Q\frac{1}{E-Q{\mathcal H}Q}Q{\mathcal H}P.
\label{Heff0}
\end{equation}
An immediate consequence of Eq. (\ref{Heff0}) is that in order 
to obtain an energy independent Hamiltonian in the model
space, it is sufficient to impose \textit{one} of the following
decoupling conditions
\begin{equation}
Q{\mathcal{H}}P=0,\label{decouplQP}
\end{equation}
or
\begin{equation}
P{\mathcal{H}}Q=0\label{decouplPQ}
\end{equation}
We note, however, that the former condition also ensures that the
$Q$-space component of the wave function $|\Psi\rangle$ vanishes,
although this is not true for its complementary left eigenstate.
Moreover, as it will become clear in the derivation of the effective
operators below, \textit{both} conditions have to be satisfied so that
one obtains energy-independent effective operators corresponding to
other observables besides the Hamiltonian.

In the case of general operators, $O$, properly transformed by the
same transformation operator $X$, e.g., the Hamiltonian $H$ in Eq.
(\ref{transfHam}), one has to compute a matrix element of the form
$\langle \tilde \Phi |{\mathcal O}|\Psi\rangle$, where $\langle \tilde
\Phi|$ corresponds possibly to another left eigenvector of
${\mathcal H}$.
Using the fact that the $Q$-component of the left eigenstate
$\langle\tilde \Phi|$ can be written similarly to Eq. (\ref{QrightEV})
\begin{equation}
\langle\tilde\Phi|Q=\langle\tilde\Phi|P{\mathcal H} Q
\frac{1}{E_\Phi-Q{\mathcal H}Q},
\label{QleftEV}
\end{equation}
one can extract the expression for the effective operator in the
model space $P$
\begin{eqnarray}
\lefteqn{{\mathcal O}_{eff}=P{\mathcal O}P+P{\mathcal H}Q
\frac{1}{E_\Phi-Q{\mathcal H}Q}Q{\mathcal O}P
+P{\mathcal O}Q \frac{1}{E_\Psi-Q{\mathcal H}Q}Q{\cal H}P\nonumber} \\
& &
+P{\mathcal H}Q\frac{1}{E_\Phi-Q{\mathcal H}Q}Q{\mathcal O}Q
\frac{1}{E_\Psi-Q{\mathcal H}Q}Q{\cal H}P.
\end{eqnarray}
As advertised, in order to obtain an energy-independent expression
for a general effective operator one needs to construct the
transformation operator $X$ so that \textit{both} decoupling 
conditions (\ref{decouplQP}) and (\ref{decouplPQ}) are satisfied.
Consequently, both left and right $P$ eigenstates of the
transformed Hamiltonian ${\mathcal H}$ have components only in the
model space. A number of other subtleties exist within this effective
operator approach \cite{ViazVary}. 

In order to determine the transformation $X$, we consider the following ansatz
\cite{Navratil:1993plb,Navratil:1993NPA}
\begin{equation}
  X=\exp(-\omega)\exp(\Omega),
 \label{ansatz}
\end{equation}
with the new operators fulfilling the additional requrements
\[
\omega=Q\omega P,
\]
\[
\Omega=P\Omega Q.
\]
Hence, the decoupling condition (\ref{decouplQP}) 
transforms into a quadratic equation for
$\omega$
\begin{equation}
  Q{\mathcal H}P=QHP-Q\omega H P +Q H\omega P -\omega H\omega=0,
  \label{eqW}
\end{equation}
which does not depend on $\Omega$,
while the decoupling condition (\ref{decouplPQ}) becomes a linear 
equation for $\Omega$ 
\begin{equation}
  P{\mathcal H}Q=PHQ+P\Omega H Q-Q\Omega\omega HQ -PH\Omega Q- PH\omega\Omega Q=0.
\end{equation}
The result of applying such a transformation is the following expressions
for the effective Hamiltonian
\begin{equation}
  {\mathcal H}_{eff}=PHP+PH\omega,
\end{equation}
which is manifestly non-Hermitian, even if the original Hamiltonian
$H$ is Hermitian, and for general effective operators:
\begin{equation}
  {\mathcal O}_{eff}=(P+\Omega-\Omega\omega)O(P+\omega),
\end{equation}
which also changes symmetry properties under the Hermitian conjugation operation.

We have made no assumption up to now about the original Hamiltonian, but in
most cases of interest, $H$ is Hermitian. For such applications, one 
can introduce an additional transformation \cite{Scholtz:1992},
so that the effective Hamiltonian in the model space is also 
Hermitian \cite{Navratil:1993plb}
\begin{equation}
  {\mathcal H}_{eff}=\frac{P+\omega^\dagger}{\sqrt{P+\omega^\dagger\omega}}
                H\frac{P+\omega}{\sqrt{P+\omega^\dagger\omega}}.
\label{effHam}
\end{equation}
Moreover, for Hermitian Hamiltonians one finds $\Omega=(P+\omega^\dagger
\omega)^{-1}\omega^\dagger$ \cite{Navratil:1993plb}, so that a general 
effective operator can also be written similarly to the 
effective Hamiltonian, i.e., involving only the operator $\omega$
\begin{equation}
  {\mathcal O}_{eff}=\frac{P+\omega^\dagger}{\sqrt{P+\omega^\dagger\omega}}
                O\frac{P+\omega}{\sqrt{P+\omega^\dagger\omega}}.
\label{effOp}
\end{equation}

There are two iterative solutions of Eq. (\ref{eqW}) that determine the
transformation operator $\omega$: one that converges to the states with the
largest  $P$-space components and is equivalent to the solution of
Krenciglowa and Kuo \cite{Krenciglowa}, and another which converges to states lying
closest to a chosen parameter appearing in the iteration procedure 
\cite{Suzuki:1980,Suzuki:1982}. However, we present here a more efficient
method to find $\omega$. It relies on the fact that the components of the
exact eigenvectors in the complementary space are mapped into the model
space. Thus, a simple and efficient means to compute the matrix elements
of $\omega$ is \cite{Navratil:2000gs}
\begin{equation}
\langle\alpha_{Q}|\omega|\alpha_{P}\rangle=
\sum_{k\in{\mathcal{K}}}\langle\alpha_{Q}|\Psi_k\rangle
\langle{\Psi_k}|\alpha_{P}\rangle^{-1},\label{omega}
\end{equation}
where $|\alpha_{P}\rangle$ and $|\alpha_{Q}\rangle$ are the basis states
of the $P$ and $Q$ spaces, respectively, and $|\Psi_k\rangle$ denotes states
from a selected set ${\mathcal{K}}$ of \textit{exact} eigenvectors
of the Hamiltonian in the full space.
The dimension of the subspace ${\mathcal{K}}$ is equal
with the dimension of the model space $P$. In the next subsection, we
will present a practical implementation of Eq. (\ref{omega}).

To conclude this brief review of the formal effective operator theory,
we would like to reiterate the main idea: in order to obtain
energy-independent operators in a restricted model space, it is
sufficient to design a transformation $X$ so that all the matrix
elements of the transformed Hamiltonian connecting the model
and the excluded space are identically zero, i.e., Eqs. 
(\ref{decouplQP}) and (\ref{decouplPQ}) are simultaneously satisfied.
Making the ansatz in Eq. (\ref{ansatz}), one can find equations which
determine the transformation, so that the decoupling conditions are satisfied.
Finally, in the case of Hermitian Hamiltonians, such as the many-body
nuclear Hamiltonain, we gave the general expressions for the
effective Hamiltonian and effective operators in the model space. 
Even in this case, one can, in principle, obtain non-Hermitian effective
Hamiltonains, but one can always make an additional transformation to
obtain a Hermitian structure, which is much more convenient to apply
to the description of a system of $A$ nucleons using realistic interactions.

\subsection{Application to the nuclear Hamiltonian}

We assume that the system of $A$ nucleons is described by the non-relativistic
intrinsic Hamiltonian
\[
 H_A=\frac{1}{A}\sum_{i<j=1}^{A}\frac{p_{ij}^{2}}{2m}
   +\sum_{i<j=1}^AV_{ij}^{NN},
\]
where $\vec{p}_{ij}=\vec{p}_{i}-\vec{p}_{j}$ are the relative momenta
between two nucleons, and $V_{ij}^{NN}$ the NN potential,
such as the local Argonne $v_{18}$ \cite{Wiringa:1995,Pieper:2001AR} or the non-local charge
dependent Bonn potential \cite{Machleidt:1995km}, which describe with high accuracy
the experimental two-nucleon data. The generalization
to include three-body forces is straightforward, but much more involved
(see, e.g., Ref. \cite{Navratil:2003ef}).
Thus, for the purpose of this paper, we neglect three-body
forces. 

In the NCSM approach, the single-particle wave functions are 
described using harmonic oscillator (HO) states. One then constructs
many-body states using a restricted set of one-body HO states.
The model space
is determined by the requirement that the
the many-body basis states can have up to $N_{max}\hbar\Omega$ excitations above the
minimum energy configuration, where $\hbar\Omega$ is the HO energy
parameter and $N_{max}$ is an integer. Including all states up to a
given HO energy allows us to separate exactly by projection containing
spurious center-of-mass (CM) motion, even when we work in a non-translationally
invariant basis.

As seen explicitly in Eq. (\ref{omega}), the solution of the $A$-body problem
is required in order to solve for the transformation operator $\omega$.
However, the eigenvectors $|\Psi_k\rangle$ are, in principle,
the final goal, as they allow computation of any properties of the system.
To practically implement the method to solve many-body problems,
we introduce the cluster approximation. This consists in finding $\omega$
for the $a$-body problem, $a<A$, and then using the effective interaction
thus obtained for solving the $A$-body system. There
are two limiting cases of the cluster approximation: 
first, when $a\rightarrow A$,
the solution becomes exact; a higher-order cluster is a better approximation
and was shown to increase the rate of convergence 
\cite{Navratil:2003ef,Navratil:2003prl}.
Second, when $P\rightarrow1$, the effective interaction approaches
the bare interaction; as a result, the cluster approximation effects
can be minimized by increasing as much as possible the size of the
model-space size.

We emphasize that in the $a$-body cluster approximation the explicit
decoupling conditions in Eqs. (\ref{decouplQP})--(\ref{decouplPQ})
are now fulfilled only for the $a$-body problem: 
\[
Q_{a}{\mathcal{H}}^{(a)}P_{a}=Q_{a}X_a
{H}_a X_a^{-1}P_a=0,
\]
where $P_{a}$, $Q_{a}$ refer to the corresponding projection operators
for the $a$-particle system. Conditions 
(\ref{decouplQP})--(\ref{decouplPQ}) are, in general, violated for
the $A$-body problem, but the errors become smaller with increasing
the size of the model space.

The rate of convergence for a fixed cluster approximation can be improved
by adding to $H_A$ a CM Hamiltonian, which also provides a single-particle
HO basis for performing numerical calculations. Doing this, we obtain

\begin{eqnarray}
\lefteqn{H_{A}^{\Omega}=H_{A}+\frac{\vec{P}^{2}}{2mA}+
\frac{1}{2}mA\Omega^{2}R^{2}}
\nonumber \\
&  & 
=\sum_{i=1}^{A}\left[\frac{\vec{p}_{i}^{2}}{2m}+
\frac{1}{2}m\Omega^{2}\vec{r}_{i}^{2}\right]
 +\sum_{i>j=1}^{A}\left[V_{ij}^{NN}-\frac{m\Omega^{2}}
{2A}(\vec{r}_{i}-\vec{r}_{j})^{2}\right]\nonumber \\
 &  & =\sum_{i=1}^{A}h_{i}+\sum_{i>j=1}^{A}v_{ij}\;,
\label{intrCM}
\end{eqnarray}
In a $a$-body cluster approximation, this ensures a dependence
of the transformation, and, therefore, of the effective
interaction on $A$. The CM term does not introduce any net influence on the converged
intrinsic properties of the many-body calculation, as we subtract
it in the final many-body calculation. Moreover, although this
addition and subtraction does not affect our exact treatment of the
CM motion, this procedure introduces a pseudo-dependence upon the
HO energy $\hbar\Omega$, and the two-body cluster approximation described
above will exhibit this dependence. In the largest model spaces, however,
important observables manifest a considerable independence of the
energy $\hbar\Omega$ and the model space size, i.e., the value of $N_{max}$.

Finally, note that even if the original Hamiltonian contained just one- and
two-body terms, the operator $X$, the transformed Hamiltonian
${\mathcal{H}}_{eff}$ [by means of Eq. (\ref{effHam})] and
transformed operators [by means of Eq. (\ref{effOp})]  all contain up to 
irreducible $a$-body terms. (The exact effective operators 
contain up to irreducible $A$-body terms.)

\section{Application to effective interactions}
\label{effInt}

The first application of the effective operator theory in the context of the
NCSM is to compute an effective interaction in a restricted model space. While
the cluster approximation described in Sec. \ref{thAp} is general for $a$
nucleons, we are currently limited by the complexity of the calculations 
to $a\leq3$. 

\begin{figure}
\centering{\includegraphics*[scale=0.55]{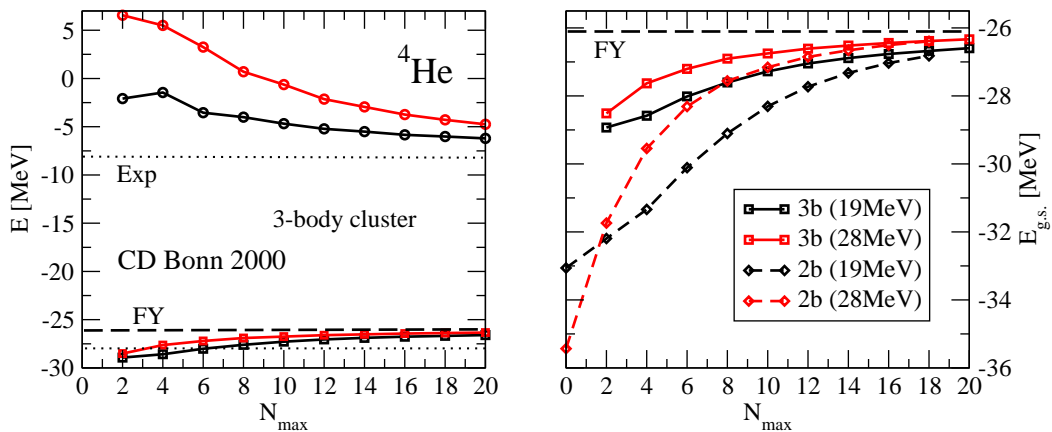}}
\caption{$^4$He: dependence of the
ground- and excited-state energies on $N_{max}$ in the three-body
cluster approximation (left panel), and comparison of the convergence
rates of the ground-state energy for the two- 
and three-body cluster approximations
(right panel). Two different HO energies (19 and 28 MeV) 
have been used in each case. The dashed line is the exact ground-state
energy \cite{gsHe4energy} for the CD Bonn 
potential used in this investigation,
while the dotted lines represent the experimental 
ground- and first excited-state energies.}
\label{he4En}
\end{figure}

In Figure \ref{he4En} we present the results for $^4$He, using both the 
two- and three-body cluster approximations. 
In the left panel, we show both the ground-
and excited-state energies using 
HO energies of 19 and 28 MeV and a
three-body cluster in order to compute the effective interaction for
$^4$He, starting from the CD Bonn interaction \cite{Machleidt:1995km}.
The convergence pattern shows a dependence upon the HO energy. 
Thus, the ground-state energy converges
faster when $\hbar\Omega=28$ MeV, but both HO energies eventually 
converge to the exact result obtained by solving, e.g., the Fadeev-Yakubovski
equations \cite{gsHe4energy}. The complete convergence of
the ground-state energy can be obtained within the NCSM, as demonstrated,
e.g., in Fig. 1 of Ref. \cite{Navratil:2005}. Because we neglect three-body
interactions, the converged result misses by a few MeV the
experimental value. Unlike for the ground state, 
the first $0^+$ excited-state energy has a faster convergence rate for 
$\hbar\Omega=19$ MeV. However, this state converges much more slowly than the
ground state, and even in the largest model spaces the results are
quite sensitive to the choice of the HO energy parameter.


As expected, a higher-body cluster approximation 
includes more correlations in the interactions,
and the convergence is faster. This is illustrated in the right panel
of Fig. \ref{he4En}, where we plot the ground-state 
energy dependence on $N_{max}$
obtained by computing the effective interaction using both the two- and 
three-body cluster approximations. The rate of convergence
is faster in the three-body cluster approximation for both HO energies
chosen for this example. 

We have used $^4$He in this section to 
illustrate the convergence properties of 
effective interactions, but the method has been applied successfully to the
description of the spectra of $p$ shell nuclei
\cite{Navratil:2000gs,Navratil:2003ef,Navratil:2000ww,Navratil:2001,Forssen:2004dk}
and beyond.

\section{Application to general operators}
\label{genEffOp}

So far, most applications of the NCSM approach have been to calculating
the effective interaction, and only a few
of publications \cite{Navratil:1996jq,Stetcu:2004bk,Stetcu:2004wh,Stetcu:2006zn} 
have investigated the renormalization of general operators 
in realistic calculations of nuclear properties. 
In Ref. \cite{Navratil:1996jq} Navr\'atil \textit{et al.}
performed large-basis NCSM calculations, which were later 
explicitly truncated into
a $0\hbar\Omega$ space and fitted to one-body quadrupole operators.
By construction these calculations contained all correlations up to
six-body due to the truncation and, hence, yielded the large effective
charge renormalizations of $1.5\; e$ for protons and $0.5\; e$ for neutrons
found empirically.
However, the full space renormalization of selected 
electromagnetic operators has been
reported only relatively recently \cite{Stetcu:2004wh,Stetcu:2006zn}.
We review below the results for one- and two-body operators.

\subsection{One-body operators}

In the $a$-body cluster approximation, the effective operators 
corresponding to an one-body operator will have, in general, 
irreducible $a$-body terms. The simplest approximation
is the two-body cluster. In order to apply it, one has 
to rewrite the original one-body contributions as a sum of 
two-body terms. For details on this procedure, we refer the reader
to, e.g., Refs. \cite{Stetcu:2004bk,Stetcu:2004wh}.

In the case of the quadrupole operator, we follow the 
procedure described in Ref. 
\cite{Stetcu:2004wh}. Selected $B(E2)$ results, obtained using the two-body
cluster approximation for $^6$Li and $^{12,14}$C are presented
in Table \ref{table1}. We have performed calculations with effective operators
only in small model spaces for several reasons. First, as expected from 
the convergence properties of effective operators mentioned in 
Sec. \ref{thAp}, larger renormalization effects are expected 
in smaller model spaces. Second, the application of the procedure
for tensor operators is much more involved, since they can 
connect states with different angular momentum or/and isospin. 
Hence, in Eq. (\ref{effOp}) one can have different
transformation operators $\omega$ to the left and to 
the right of the bare operator. Moreover, the number of 
two-body matrix elements for non-scalar operators 
can be orders of magnitude larger than the number of 
one-body matrix elements. Finally,
the main purpose of these investigations was a qualitative 
understanding of the influence of effective operators and not a highly 
accurate description of the experimental data.

\begin{table}
\caption{$B(E2)$, in $e^2 fm^4$, for selected nuclei and model 
spaces, using the bare operator and the effective operator, computed in the
two-body cluster approximation.}
\label{table1}
\begin{indented}
\lineup
\item[]\begin{tabular}{@{}*{5}{l}}
\br
Nucleus & Observable & Model Space & Bare operator & Effective operator\cr
\mr
$^6$Li  & $B(E2,1^+0\rightarrow 3^+0)$& \0$2\hbar \Omega$ & \02.647 & 2.784 \cr 
$^6$Li  & $B(E2,1^+0\rightarrow 3^+0)$& $10\hbar \Omega$ &  10.221 &-- \cr
$^6$Li  & $B(E2,2^+0\rightarrow 1^+0)$ & \0$2\hbar \Omega$ &\02.183 & 2.269\cr
$^6$Li  & $B(E2,2^+0\rightarrow 1^+0)$ & $10\hbar \Omega$& \04.502 &-- \cr
$^{10}$C & $B(E2, 2_1^+ 0\rightarrow 0^+0)$ & \0$4\hbar \Omega$ & \03.05 & 3.08 \cr
$^{12}$C & $B(E2, 2_1^+ 0\rightarrow 0^+0)$ & \0$4\hbar \Omega$ & \04.03 & 4.05 \cr
\br
\end{tabular}
\end{indented}
\end{table}

As illustrated in Table \ref{table1}, the effective operators 
have very little effect on the results for the qudrupole 
transitions. For $^6$Li, we also present the $B(E2)$ values 
obtained in $10\hbar\Omega$ model space \cite{Navratil:2001}.
If the effect of the renormalization of the quadrupole operator 
had been significant, then the
$B(E2)$ values in the small model spaces would be closer to the results in the
$10\hbar\Omega$ model space, which is obviously not the case.
The same weak renormalization
can be observed for the carbon isotopes, listed in Table \ref{table1}.
This is contrary to the previous
investigation in the framework of the NCSM \cite{Navratil:1996jq},
which reported 
obtaining the correct effective proton and neutron phenomenological charges.
However, the main difference is that the $^6$Li calculation in 
Ref. \cite{Navratil:1996jq} included up to six-body correlations.
Comparison of the two results
suggests that higher-order clusters can play an important role 
in the renormalization of the quadrupole operator.

\subsection{Two-body operators}

In a previous publication \cite{Stetcu:2004wh}, we used a 
two-body Gaussian operator to demonstrate the dependence of 
the renormalization upon the range of the operator. In a recent
paper \cite{Stetcu:2006zn}, we computed the 
longitudinal-longitudinal distribution
function, part of the inclusive $(e,e')$ response. In this paper we
present similar results, obtained in smaller model
spaces but converged nevertheless at high momentum transfer 
(=short range), because we use the appropriate
effective operators. Moreover, the effect of the renormalization is larger
in smaller model spaces, as noted before.

To define the longitudinal-longitudinal distribution function,
one starts with the Coulomb sum rule
\begin{equation}
S_{L}(q)=\frac{1}{Z}\int_{\omega_{el}}^{\infty}
d\omega S_{L}(q,\omega)\label{eq:CSR},
\end{equation}
which is the total integrated strength measured in electron scattering.
In Eq. (\ref{eq:CSR}), $S_{L}(q,\omega)=R(q,\omega)/|G_{E,p}(q,\omega)|^{2}$,
with $R(q,\omega)$ the longitudinal response function and $G_{E,p}(q,\omega)$
the proton electric form factor, while $\omega_{el}$ is the energy
of the recoiling $A$-nucleon system with $Z$ protons. 
$S_{L}(q),$ which is related to the Fourier transform of
the proton-proton distribution function \cite{Drell:1958,McVoy:1962},
can be expressed in terms of the longitudinal form factor $F_L(q)$
and the longitudinal-longitudinal distribution function $\rho_{LL}$ 
as \cite{Schiavilla:1993tk}

\begin{eqnarray*}
S_{L}(q) & = & \frac{1}{Z}\langle g.s.|\rho_{L}^{\dagger}
(\mathbf{q})\rho_{L}(\mathbf{q})|g.s.
\rangle-\frac{1}{Z}|\langle g.s.|\rho_{L}(\mathbf{q})|g.s.\rangle|^{2}\\
 &  & \equiv1+\rho_{LL}(q)-ZF_{L}(q)/G_{E,p}(q,\omega_{el}).
\end{eqnarray*}
If one neglects relativistic corrections and 
two-body currents, then $\rho_{L}(\mathbf{q})$ is the charge operator,

\[
\rho_{L}(\mathbf{q})=\frac{1}{2}\sum_{i=1}^{A}
\exp(i\mathbf{q}\cdot\mathbf{r}_{i})(1+\tau_{z,i}),\]
so that the longitudinal-longitudinal distribution function
becomes \cite{Schiavilla:1993tk}
\[
\rho_{LL}(q)=\frac{1}{4Z}\sum_{i\neq j}
\langle g.s.|j_{0}(q|\mathbf{r}_{i}-\mathbf{r}_{j}|)
(1+\tau_{z,i})(1+\tau_{z,j})|g.s.\rangle.
\]

\begin{figure}
\centering{\includegraphics*[scale=0.8]{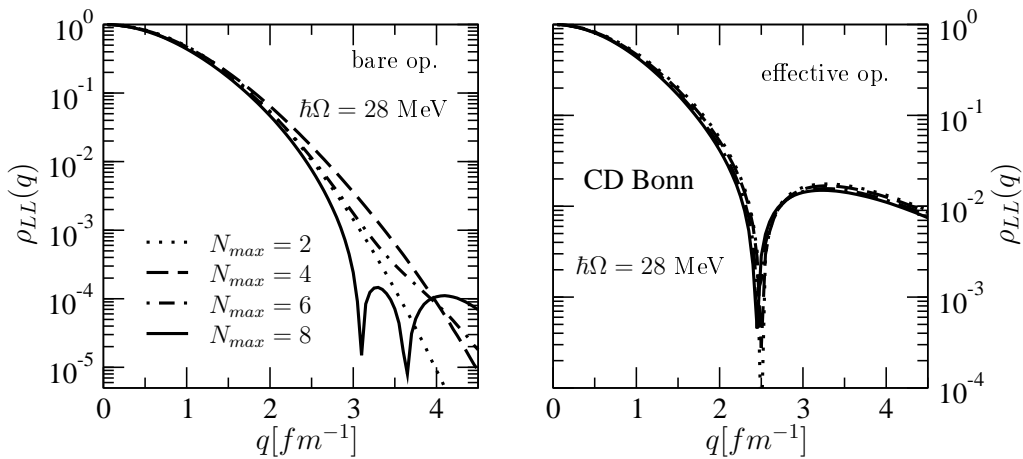}}
\caption{The longitudinal-longitudinal distribution function
in $^4$He, obtained using
bare operators (left panel) and effective operators (right panel).
The HO energy used in this calculation
was $\hbar\Omega=28$ MeV, while the NN interaction was CD Bonn.}
\label{he4rhoLL}
\end{figure}

In Figure \ref{he4rhoLL} we present the results for the 
longitudinal-longitudinal distribution function for $^4$He. 
At high momentum transfer, the results obtained 
using bare operators depend strongly upon the model space. On
the other hand, the results obtained with effective
operators are model space invariant at high $q$, although Figure
\ref{he4En} shows that the wave function is not fully converged, since
the energy is not converged in these very small model spaces. They
agree with the values computed in larger model spaces
and different HO energies given in Ref. \cite{Stetcu:2006zn}.
At intermediate momentum transfer, i.e., $q\approx 2.5$ fm$^{-1}$,
even the effective operator results vary. This
effect is due to the fact that the long range part of the operator
has not yet converged in these small model spaces.

Similar results for the longitudinal-longitudinal distribution
function have been obtained for $^{12}$C, where calculations
in very large model spaces are not possible. However, even in the
smallest model space, $0\hbar\Omega$, we were able to obtain
good results for high momentum transfer, which reproduce the
values in larger model spaces \cite{Stetcu:2006zn}.

As demonstrated in Ref. \cite{Stetcu:2004wh} with a 
two-body Gaussian operator and illustrated here for the
longitudinal-longitudinal distribution function, in the two-body
cluster approximation the renormalization depends strongly of
the range of the operator. Short range operators (high momentum
transfer) are very well renormalized and the results become 
model-space independent even in the two-body cluster approximation,
while long-range operators, such as
the quadrupole transition operator, or the longitudinal-longitudinal
distribution function for small and intermediate momentum
transfer, are only weakly renormalized.

\section{Conclusions}
\label{concl}

In this paper, we have reviewed the application of the effective operator
theory in the framework of the NCSM. While in the derivation one can
obtain non-Hermitian operators that are more suitable for some
applications \cite{Bender:2005sc},
we construct, by means of additional transformations, Hermitian operators,
which are easier to utilize in large scale calculations.

The \textit{ab initio} NCSM has been applied successfully 
to the description of the nuclear spectra for light nuclei 
\cite{Navratil:2000gs,Navratil:2003ef,Navratil:2000ww,Navratil:2001,Forssen:2004dk},
i.e., $A\leq 16$, and beyond \cite{ne20NCSM,A48}. The wave functions obtained
can be used to calculate and predict nuclear properties, 
such as the proton radii of
halo nuclei \cite{Caurier:2005rb}, or the astrophysical $S$-factor
\cite{Navratil:2006tt}, to cite just a couple of the most recent results.
Moreover, for light nuclei, the precision of the NCSM method makes it
possible  to investigate the reliability of the chiral 
nuclear interaction. This follows from the fact that the properties, 
e.g., energy spectra, of $p$-shell
nuclei are sensitive to the subleading parts of the chiral interactions, 
including three-nucleon forces \cite{Nogga:2005hp}.
For heavier nuclei, another approach, designed to improve the 
convergence of the results, has been recently
proposed, combining the inverse $J$-matrix scattering technique and
the NCSM \cite{Shirokov:2005bk}.

In the two-body cluster approximation, one has now the ability to compute not
only the effective interaction, but also the consistent
effective operators corresponding
to scalar and tensor observables. We have shown a strong dependence on the
renormalization of the range of the bare operator. Thus, if the
operator is of short range, then one obtains a good renormalization
in the two-body
cluster approximation, as the unitary transformation used to obtain the
effective interaction renormalizes mostly the short-range repulsion 
of the potential. Consequently, 
one obtains model-space independence results for such observables. 
Long-range operators, on the other hand, are only weakly renormalized
at the two-body cluster level.
In order to accomodate the long-range correlations one has to increase
the model space and/or use a higher-order cluster approximation.
The success of latter was demonstrated
by the good results for the $0\hbar\Omega$ effective charges obtained in a
restricted NCSM calculation \cite{Navratil:1996jq}.

\section*{Acknowledgments}
I.S. and B.R.B acknowledge partial support by NFS grants PHY0070858
and PHY0244389. The work was performed in part under the auspices
of the U. S. Department of Energy by the University of California,
Lawrence Livermore National Laboratory under contract No. W-7405-Eng-48.
P.N. received support from LDRD contract 04-ERD-058. J.P.V. acknowledges
partial support by USDOE grant No DE-FG-02-87ER-40371.\\


\end{document}